\documentclass[preprint,            
               showpacs,longbibliography, 
               showkeys,preprintnumbers,
               aps, prd,
               letterpaper,
               superscriptaddress,
               nofootinbib,
               tightenlines,
               floats,floatfix
               ]{revtex4}

\usepackage{physics}
\usepackage{epsf}
\usepackage{graphicx,color}
\usepackage{subfigure}
\usepackage{latexsym}
\usepackage{amsmath,amssymb}        
\usepackage[colorlinks=true,linkcolor=blue,citecolor=blue]{hyperref}
\usepackage{cleveref}
\usepackage{mathrsfs}
\usepackage{comment}
\usepackage{soul}
\usepackage{cancel}
\usepackage{placeins}

\usepackage[top=2.5cm,bottom=2.5cm,left=2.5cm,right=2.5cm]{geometry}
\newcommand\myemail[1]{{\color{red}  #1}}
\begin{document}

\begin{center}
  \large
  \textbf{Construction of ground state solutions of the Gross-Pitaevskii-Poisson system using genetic algorithms}\\[1em]
  \small
Carlos Tena-Contreras\textsuperscript{$\dag$}\footnote{\myemail{1423327c@umich.mx}}, 
Francisco S. Guzmán\textsuperscript{$\dag$}\footnote{\myemail{francisco.s.guzman@umich.mx}}, 
Iván Álvarez-Ríos\textsuperscript{$\dag$}\footnote{\myemail{ivan.alvarez@umich.mx}}\\[0.5em]
  \textsuperscript{$\dag$}\textit{Instituto de Física y Matemáticas, Universidad Michoacana de San Nicolás de Hidalgo,\\ Edificio C-3, Cd. Universitaria, 58040 Morelia, Michoacán, México}\\[1em]
  \today
\end{center}

We present the construction of the ground state of the Gross-Pitaevskii-Poisson equations using genetic algorithms. By employing numerical solutions, we develop an empirical formula for the density that works within the considered parameter space. Through the analysis of both numerical and empirical solutions, we investigate the stability of these ground state solutions. Our findings reveal that while the numerical solution outperforms the empirical formula, both solutions lead to similar oscillation modes. We observe that the stability of the solutions depends on specific values of the central density and the nonlinear self-interaction term, and establish an empirical criterion delineating the conditions under which the solutions exhibit stability or instability.

\vspace{0.5cm}

\vspace{0.5cm}

\section{Introduction}

The study of solitonic cores in physical systems has attracted significant interest due to their relevance in various fields, including cosmology, astrophysics, and condensed matter physics. These localized, stable structures arise from the balance between attractive and repulsive forces, leading to unique properties and behaviors. 

One specific field where this system has become particularly interesting is that of bosonic dark matter. This model assumes that dark matter particle is a spinless ultralight boson with mass of order $10^{-19}-10^{-23}\rm{eV/c}^2$ and there are various reviews that draw the a general panorama of the subject \cite{Suarez:2013,Hui:2021tkt,ElisaFerreira,Chavanis2015,Niemeyer_2020}. 
The dynamics of this type of matter in the mean field approximation is ruled by the Gross-Pitaevskii-Poisson (GPP) system of equations and the system can be seen as a Bose gas trapped by the gravitational field sourced by itself \cite{GuzmanUrena2006}. When the gas has no self-interaction it is called the Fuzzy Dark Matter (FDM) regime \cite{Hu:2000}, and is the main stream of the subject that has led to study both local (e.g.\cite{Schwabe:2016,Mocz:2017wlg}) and structure formation dynamics \cite{Schive:2014dra,Veltmaat_2018,Schive:2014hza}.

The interest of this model is that for this ultralight particle the de Broglie wavelength is large and thus the structures have a minimum size. In fact, it was found since the breakthrough simulation in \cite{Schive:2014dra}, that cores are an essential part of structures, they are an attractor profile surrounded by an envelope with high kinetic energy. It resulted that these cores could be fitted by am empirical density profile that coincides numerically with the solution of the ground state solutions of the GPP system \cite{GuzmanUrena2004}. Ever since, these cores have been considered literally the keystone of structures in FDM. Bounds on the values of self-interactions arise from local and cosmological constraints 
\cite[see e.g.][]{RindlerDaller:2013,Desjacques:2017fmf,Cembranos:2018ulm,Chavanis:2020rdo,Delgado:2022vnt}.

The generalization to include self-interacting bosons is a natural extension, and limits to the self-interaction regime is a matter of interest, because the dark matter distribution differs from that of FDM \cite{Chavanis_2023}, time scales of saturation and relaxation change \cite{ChengNiemeyer2021} and ultimately there is a Tomas-Fermi regime with rather simple density profiles \cite{Boehmer:2007}.

This background leads us to revisit some properties and stability of solitonic cores with self-interaction that will show useful. We solve the well known eigenvalue problem for ground state solutions of the GPP system, with a rather unusual but efficient method that involves Genetic Algorithms. We then propose an empirical formula that describes its profile. We then study the stability of the solutions and those associated to the empirical formula. We compare the reaction of both to truncation error perturbations in both amplitude and frequencies of the oscillation modes triggered. We point out to interesting differences and potential implications within the dark matter context.


\section{Model and Equations}
\label{sec:model}

In a Bose-Einstein Condensate (BEC), a significant portion of the particles occupy the lowest quantum state, resulting in overlapping wave functions. Consequently, the state of a BEC can be effectively described by a collective wave function, known as the order parameter, $\Psi(t, \vec{x})$. Due to interactions between bosons, which induce nonlinear effects, this collective wave function satisfies the Gross-Pitaevskii (GP) equation:

\begin{equation}
\mathrm{i}\hbar \frac{\partial \Psi}{\partial t} = \left[ -\frac{\hbar^2}{2m} \nabla^2 + mV + g|\Psi|^2 \right] \Psi, \label{eq:GP}
\end{equation}

\noindent where $\hbar$ is the reduced Planck constant, $m$ is the mass of the boson, $V$ is the external potential acting as a gas trap,  and $g = \frac{4\pi \hbar^2 a_s}{m}$ is the nonlinear coefficient, where $a_s$ is the scattering length of two interacting bosons.

The concept of BEC Dark Matter (BEC-DM) hypothesizes that dark matter exists in the form of a BEC, where the trapping potential is self-generated by the bosonic ensemble through the Poisson equation:

\begin{equation}
\nabla^2 V = 4\pi G m |\Psi|^2.
\label{eq:Poisson}
\end{equation}

\noindent The nonlinear system (\ref{eq:GP}-\ref{eq:Poisson}) is known as the Gross-Pitaevskii-Poisson (GPP) system and exhibits scale invariance, described by the transformation \cite{GuzmanUrena2004}:

\begin{equation}
    \left\lbrace t, \vec{x}, \Psi, V, g \right\rbrace \to \left\lbrace \lambda^{-2} t, \lambda^{-1} \vec{x}, \lambda^2 \Psi, \lambda^2 V, \lambda^{-2} g \right\rbrace,
    \label{eq:lambda-invariant}
\end{equation}

\noindent where $\lambda$ is a scaling factor.

An alternative description of the GPP system (\ref{eq:GP}-\ref{eq:Poisson}) can be obtained through the Madelung transformation $\Psi = \sqrt{\rho/m} e^{iS/\hbar}$, where $\rho = m |\Psi|^2$ is the mass density, and $S$ represents the phase of the wave function. By separating the real and imaginary parts and defining the fluid velocity as $\vec{v} = \nabla S / m$, it is possible to rewrite the GPP system as \cite{ShapiroCreTail,AlvarezGuzmanMadelung}:

\begin{eqnarray}
    \frac{\partial \rho}{\partial t}  + \nabla\cdot(\rho\vec{v}) & = & 0,
    \label{eq:mass conservation} \\
    \frac{\partial (\rho \vec{v})}{\partial t} + \nabla\cdot(\rho \vec{v}\otimes\vec{v} + p_{\text{SI}} I) & = & -\rho\nabla(Q + V),
    \label{eq:momentum conservation} \\
    \nabla^2 V & = &  4\pi G \rho,
    \label{eq:Poisson Madelung}
\end{eqnarray}

\noindent where $p_{\text{SI}} = \frac{g}{2}\rho^2$ is the self-interaction pressure, and $Q = -\frac{\hbar^2}{2 m} \frac{\nabla^2 \sqrt{\rho}}{\sqrt{\rho}}$ is the quantum potential. In this framework, the GPP system is known as the quantum hydrodynamic formulation. 

If we take the limit $\hbar/m \to 0$, the quantum potential becomes zero, and the classical hydrodynamic equations are recovered with a polytropic equation of state with polytropic constant $K=g/2$ and polytropic index $n=1$. In this limit, we can see the effect of the nonlinear term in the GP equation classically, which results from two-body dispersion interactions between bosons:

\begin{itemize}
    \item Repulsive self-interaction ($g>0$): The gas has positive pressure opposing the gravitational force, allowing stable structures.
    \item Attractive self-interaction ($g<0$): No classical behavior is possible since negative pressure is not physically acceptable. 
    \item Without self-interaction ($g=0$): The limit of positive self-interaction when the polytropic constant tends to zero corresponds to a dust-like state.
\end{itemize}

\noindent When $\hbar/m > 0$, the quantum potential reintroduces the possibility of structure formation regardless of the value of $g$. The quantum potential $Q$ is crucial for this behavior. As in any macroscopic system, global quantities can be measured. Some of these are:

\begin{eqnarray}
    M_T & = & m\int |\Psi|^2 d^3x = \int \rho d^3 x, \label{eq:mass} \\
    W & = & \dfrac{m}{2}\int \Psi^* V \Psi d^3x = \dfrac{1}{2} \int \rho V d^3x, \label{eq:potenergy} \\
    K & = & -\dfrac{\hbar^2}{2 m}\int \Psi^*\nabla^2\Psi d^3x \nonumber \\ 
     & = & \frac{1}{2}\int \rho |\vec{v}|^2d^3x + \frac{1}{2}\int |\nabla\sqrt{\rho}|^2 d^3x \nonumber \\
     & = & K_v + K_\rho \label{eq:kinenergy} \\
     I & = & \dfrac{g}{2}\int |\Psi|^4 d^3x = \int p_{\text{SI}} d^3x,
\end{eqnarray}

\noindent where $M_T$ is the mass, $W$ is the potential energy, $K$ is the total kinetic energy (with $K_v$ from classical contributions and $K_\rho$ from quantum effects), and $I$ is the self-interaction energy. The total energy is defined as $E = K + W + I$, and the virial function is $Q = 2K + W + 3I$ (see e.g. \cite{Wang2001}). All these quantities are helpful for the diagnostics of any solution of the system (\ref{eq:GP}-\ref{eq:Poisson}), and here we will use them for equilibrium solutions.


\subsection{Adimensionalization of the System}

Transforming the system into a dimensionless coordinate system ensures uniformity of units and prevents issues arising from disparate scales when performing numerical calculations. To achieve this, we perform the following transformations: $t = \Tilde{t} t_0$, $\vec{x} = \Tilde{\vec{x}} x_0$, $g = \Tilde{g} g_0$, $\Psi = \Tilde{\Psi} \Psi_0$, $V = \Tilde{V} V_0$, and $\rho = \Tilde{\rho} \rho_0$, where tilde variables are dimensionless and are said to be  code units, while tilded ones are physical. Appropriate scale factors for the GPP system are:

\begin{eqnarray}
t_0 & = & \frac{m}{x_0^2 \hbar}, \nonumber \\
g_0 & = & 4\pi G m^2 x_0^2, \nonumber \\
\Psi_0 & = & \frac{\hbar}{\sqrt{4\pi G m^3} x_0^2}, \\
V_0 & = & \left(\frac{\hbar}{m x_0}\right)^2, \nonumber \\
\rho_0 & = & \frac{\hbar^2}{4\pi G m^3 x_0^4}. \nonumber
\end{eqnarray}

Thus, the system effectively possesses a single degree of freedom, which we express in terms of a scaling factor $x_0$, equivalent to the transformation (\ref{eq:lambda-invariant}). With these new variables, the GPP system can be rewritten in dimensionless units as:

\begin{equation}
\mathrm{i}\dfrac{\partial \Psi}{\partial t} = \left[-\dfrac{1}{2}\nabla^2 + V + g|\Psi|^2\right]\Psi,
\label{eq:GPadi}
\end{equation}

\begin{equation}
\nabla^2 V = |\Psi|^2,
\label{eq:Poissonadi}
\end{equation}

\noindent where we omit the tilde in all variables. Our objective now is to determine the ground state of the stationary version of the GPP system (\ref{eq:GPadi}-\ref{eq:Poissonadi}) since excited states are unstable \cite{GuzmanUrena2006}.

\subsection{The Stationary GPP System}

Stationary GPP equations are constructed by assuming spherical symmetry and that the order parameter can be rewritten as $\Psi(t, \vec{x}) = \psi(r)e^{-i\omega t}$, with $\omega$ an eigenfrquency and $\psi(r)$, a real function of the radial coordinate $r$. With these assumptions, the GPP system is written as follows, according to \cite{Ruffini:1969,GuzmanUrena2004}:

\begin{equation}
    -\frac{1}{2 r^2}\frac{d}{dr}\left(r^2\frac{d\psi}{dr}\right) + V\psi + g\psi^3 = \omega \psi,
    \label{eq:GPStationary}
\end{equation}

\begin{equation}
    \frac{1}{r^2}\frac{d}{dr}\left(r^2\frac{dV}{dr}\right) = \psi^2.
    \label{eq:PoissonStationary}
\end{equation}

\noindent To ensure physically acceptable solutions, we impose certain boundary conditions. For the stationary order parameter $\psi$, we require that $\psi(0) = \psi_c$, $\psi'(0) = 0$, and $\lim_{r \to \infty} \psi = \lim_{r \to \infty} \psi' = 0$. 

For the gravitational potential $V$, we set $V(0) = V_c$ and $V'(0) = 0$. The choice of $V_c$ can be arbitrary since shifting this condition to $V_c + V_a$ is equivalent to finding an eigenvalue $\omega + V_a$ for some arbitrary value $V_a$. These boundary conditions ensure physically meaningful solutions that satisfy the requirements of regularity and isolation.

Since this set of equations will be solved numerically, it is convenient to write it as a first-order system by defining the variables $\phi = r^2 \psi'$ and $M = r^2 V'$, where ${}' = d/dr$ is the derivative operator. The above system is then rewritten as:

\begin{eqnarray}
    \psi' & = & \frac{\phi}{r^2}, \label{eq:psir} \\
    \phi' & = & 2r^2\left(V + g\psi^2 - \omega\right)\psi, \label{eq:phir} \\
    V' & = & \frac{M}{r^2}, \label{eq:Vr} \\
    M' & = & r^2 \psi^2. \label{eq:Mr}
\end{eqnarray}

\noindent with the boundary conditions $\psi(0) = \psi_c$, $\phi(0) = 0$, $V(0) = V_c$, $M(0) = 0$, and $\lim_{r\to\infty}\psi(r) = \lim_{r\to\infty}\psi'(r) = 0$. This set of equations along with the boundary conditions define an eigenvalue problem, where the eigenvalue is $\omega$.

For ease, it is convenient to define the vector $u = (\psi, \phi, V, M)$ and the right side of the system as $f(r,u) = \left(\frac{u_2}{r^2}, 2r^2(u_3 + g u_1^2 - \omega)u_1, \frac{u_4}{r^2}, r^2 u_1^2\right)$. The system can be written compactly as:

\begin{eqnarray}
    \left\lbrace
    \begin{matrix}
    \frac{du}{dr} & = & f(r,u), \\
    u(0) & = & u_0,
    \end{matrix}
    \right.
    \label{eq:u}
\end{eqnarray}

\noindent where $u_0 = (\psi_c, 0, V_c, 0)$. 

%

\section{Numerical Methods} 
\label{sec:Numerical Methods}

Different strategies can be employed to numerically solve the systems of equations above. Some of them solve the system on a discrete domain, traditionally using a shooting routine, like in the flagship reference \cite{Ruffini:1969} and most of follow up papers. We will use a discrete domain but not a shooting method.

\subsection{Stationary system}\label{Stationary system}

We construct the solution on a finite domain $D = [0, r_{\text{max}}]$, where the boundary conditions are redefined approximately as $\psi(r_{\text{max}} ) \approx \psi'(r_{\text{max}}) \approx 0$, that is, we use a finite value $r_{max}$ in which we seek to satisfy the boundary conditions approximately at the external boundary $r_{max}$.

We define the discrete domain as $D^d = \{r_i \in D | i = 0,...,N_r\}$ where $N_r$ is the number of points. The simplest way to construct $D^d$ is by employing a uniform partition, where the points are chosen as $r_i = i \Delta r$, with $\Delta r = r_{\text{max}} / N_r$ the resolution of the discrete domain.

Note that in order to integrate the system (\ref{eq:psir}-\ref{eq:Mr}), we must set the parameters $\psi_c$, $V_c$, $g$, $\omega, r_{max}$, from which we do not know the eigenvalue $\omega$, and for reasons of numerical precision, the upper radius $r_{\text{max}}$. Therefore, it is necessary to find these values in such a way that they approximately satisfy the boundary conditions at the outer boundary $r_{\text{max}}$.

Instead of using the shooting method to search for the eigenvalue $\omega$ of the problem (\ref{eq:psir}-\ref{eq:Mr}) as traditionally (e.g. \cite{Ruffini:1969,GuzmanUrena2006}), here we propose an alternative method based on optimization.

\subsection{Description of the Eigenvalue Search Method}

We search for the eigenvalue $\omega$ that satisfies the boundary conditions at $r_{\text{max}}$. To accomplish this, we employ a genetic algorithm (GA), which is rooted in the theory of evolution. In a GA, an initial population exists within a defined environment. Each individual in this population is assigned a fitness level, representing their suitability for survival in the environment. This fitness level is determined solely by the DNA of each individual of the population.

Better-adapted individuals have a higher likelihood of reproducing and passing on their genetic material to subsequent generations. Offspring are generated from two parents, each contributing approximately 50\% of their genetic material. However, in nature, offspring may adapt better to the environment than their parents due to mutations in their DNA. This iterative process continues for many generations until significant changes are observed in the population.

Based on this understanding of evolution, we outline our GA as follows:

\begin{enumerate}
    \item \textbf{Define the problem:} In our context, each individual represents a potential solution to the eigenvalue problem of system (\ref{eq:u}). The DNA chain determining each individual is represented as $\textbf{DNA} = (\omega, r_{\text{max}})$, where the components of these vectors are called genes.
    
    \item \textbf{Initialize the population:} The population is generated randomly, with a constant size $N_{\text{population}}$ maintained throughout the evolution.
    
    \item \textbf{Fitness function:} Define the fitness function as
    \begin{eqnarray}\label{eq: fitness}
        f(\omega, r_{\text{max}}) = \left[\psi(r_{\text{max}})^2 +  \psi^\prime(r_{\text{max}})^2\right]^{-1},
    \end{eqnarray}
    \noindent where $\psi(r)$ is the solution of the system (\ref{eq:u}) associated with the value $\omega$. The choice of the form of equation (\ref{eq: fitness}) is due to the fact that both the wave function and its derivative must vanish in the limit when $r \rightarrow r_{max}$,  and we would like the violation of the condition on $\psi$ and on $\psi^{\prime}$ to be of the same order; then we define the fitness as the inverse of the mean squared violation of the separate violations of $\psi$ and on $\psi^{\prime}$.
    
    \item \textbf{Selection:} Use an elitist method to select the best $N_{\text{best}}$ individuals: The value of the fitness function of each element of the population is obtained and they are arranged in such a way that those with a higher fit are first in the list.
    
    \item \textbf{Reproduction:} Select two random parents from the first $N_{\text{best}}$ to generate a new individual. The DNA genes of the new individual are randomly selected from the genes of the parents.
    
    \item \textbf{First Mutation:} After creating a new individual, apply a mutation where each gene in the DNA chain has a probability of being amplified by a factor of 1.5.
    
    \item \textbf{Replacement:} Repeat steps 5-6 for $N_{\text{population}} - N_{\text{best}}$ to generate the remaining individuals.
    
    \item \textbf{Second Mutation:} Apply a differential mutation to the entire population. For each individual with $\textbf{DNA}_i$, select two other individuals with $\textbf{DNA}_1$ and $\textbf{DNA}_2$ randomly. Generate a new individual with $\textbf{DNA}_{\text{new}} = \textbf{DNA}_i + 0.1(\textbf{DNA}_2 - \textbf{DNA}_1)$. If $f(\textbf{DNA}_{\text{new}}) > f(\textbf{DNA}_i)$, replace the $i$th individual with the new one.
    
    \item \textbf{Stop condition:} Repeat steps 4-8 for multiple generations until $f(\textbf{DNA}) > 10^8$ for at least one individual in the population.
\end{enumerate}

By applying this algorithm, it is possible to find a solution to our problem by specifying only the amplitude of the order parameter $\psi_c$ at the origin and the coefficient of the nonlinear term $g$. Let us remember that the choice of $V_c$ can be made arbitrarily; however, once the solution is found, we can rescale the gravitational potential and the eigenvalue as follows:

\begin{eqnarray}
    \omega  - V(r_{\text{max}}) - \frac{M(r_{\text{max}})}{r_{\text{max}}} & \to & \omega, \nonumber \\
    V - V(r_{\text{max}}) - \frac{M(r_{\text{max}})}{r_{\text{max}}} & \to & V, \nonumber 
\end{eqnarray}

\noindent so that the gravitational potential satisfies monopolar boundary conditions.

Finally, the specific parameters of the GA for the solution of the eigenvalue problem are that $N_{population}=500$ whereas half of the population is selected from each generation $N_{best}=250$ is used to survive and crossover.


\section{Stationary solutions}
\label{sec:results}


We explore a parameter space that includes various  values of $g$ and $\psi_c$ within the range $(\psi_c, g) \in [1, 2] \times [-0.5, 0.5]$.

\subsection{Case $g=0$}

\begin{figure}
    \centering
    \includegraphics[width=6.5cm]{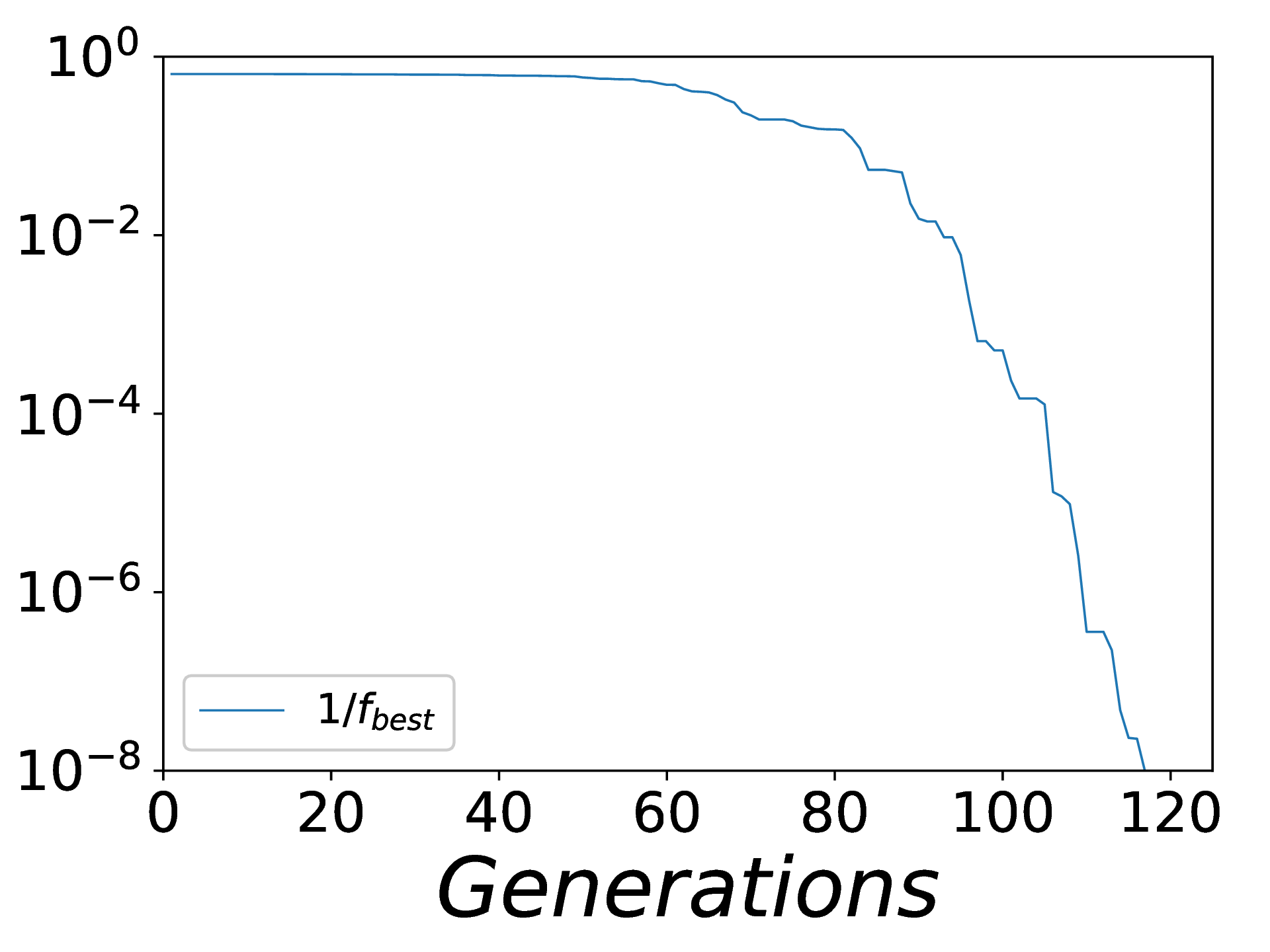}
    \includegraphics[width=7.0cm]{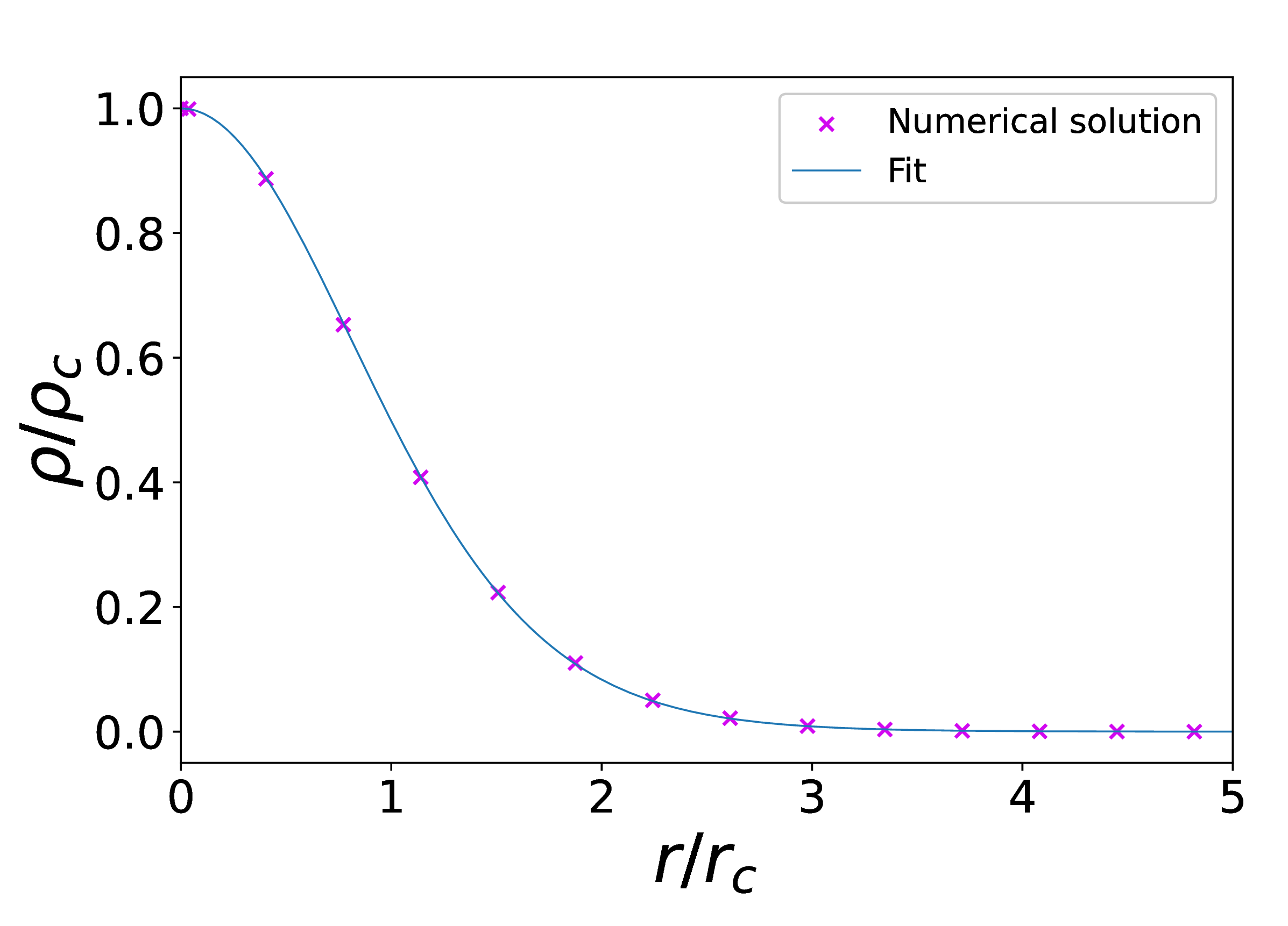}
    \caption{On the left we show the inverse of the best individual's fitness as a function of generations for the case $(\psi_c, g) = (1, 0)$, demonstrating that the GA method provides solutions approaching to the satisfaction of stationary GPP equations very rapidly. On the right we compare the solution found by the GA with the empirical formula (\ref{eq:soliton}), for the case with $\rho_c = 1$ and $r_c \approx 1.306$.}
    \label{fig:convergence}
\end{figure}

The $\lambda-$invariance  (\ref{eq:lambda-invariant}) when $g = 0$, implies that the stationary solution is uniquely determined for $\psi_c = 1$. The inverse fitness of the best individual in the genetic algorithm (GA) for $(\psi_c, g) = (1, 0)$ is depicted on the left of Figure \ref{fig:convergence}, illustrating convergence after 120 generations. The fitness reaches an optimal value of approximately $10^8$, corresponding to an error of $10^{-8}$. This tolerance is used in subsequent solutions. Once the GA finds an optimal solution, the resulting genes yield $(r_{\text{max}}, \omega) \approx (12.24, -0.6922)$, consistent with previous well known results (e.g. \cite{GuzmanUrena2004}).

The ground state for $g = 0$ can be approximated by the following empirical formula, found from structure formation simulations of Fuzzy Dark Matter \cite{Schive:2014hza}:

\begin{equation}
    \rho_{\text{core}}(r)|_{g=0} = \rho_c \left[1 + \left(2^{1/8}-1\right)\left(\frac{r}{r_c}\right)^2\right]^{-8}.
    \label{eq:soliton}
\end{equation}

\noindent which is a formula found to match the density profile of the ground state solution of the GPP system for $g=0$.
The core radius $r_c$ is defined as the radius where the density $\rho(r_c)$ is half of the central density $\rho_c$. Setting $r_c \approx 1.306$ matches Eq. \eqref{eq:soliton} with $\rho_c = 1$, specific to this case as illustrated on the right of Figure \ref{fig:convergence}. The relation between $r_c$ and $\rho_c$ is established using the $\lambda$-scaling \eqref{eq:lambda-invariant}, leading to $\rho_c \approx (1.306/r_c)^4$, for either arbitrary central density or core radius. In physical units,

\begin{equation}
    \rho_c \approx \frac{\hbar^2}{4\pi G m^2}\left(\frac{1.306}{r_c}\right)^4 \approx 1.983 \times 10^7 \left(\frac{\text{kpc}^4}{m_{22}^2 r_c^4}\right) \text{M}_\odot,
\end{equation}

\noindent with $m = m_{22} \times 10^{-22} \text{ eV}/c^2$. This formula practically relates tightly core radius and central density \cite{Mocz:2017wlg}.

\subsection{Case $g \ne 0$}

For the solutions with non-zero values of $g$ we propose a generalization of the empirical density profile given by 

\begin{equation}
    \rho_{\text{core}}(r) = \rho_c \left[1 + \left(2^{1/8}-1\right)\left(\frac{r}{r_c}\right)^{2+\beta}\right]^{-8},
    \label{eq:soliton_general}
\end{equation}

\noindent where $r_c = r_c(\rho_c, g)$ and $\beta = \beta(\rho_c, g)$ determine the core radius and exponent, respectively. The total mass of the core can be integrated to give

\begin{eqnarray}
    M_{\text{core}} & = & \int_0^\infty \rho_{\text{core}}(r)r^2 dr \nonumber \\
    & = & \frac{(2^{1/8}-1)^{-\frac{3}{2+\beta}}}{5040} \rho_c r_c^3 \frac{\Gamma\left(\frac{3}{2+\beta}\right)\Gamma\left(8-\frac{3}{2+\beta}\right)}{2+\beta} \nonumber \\
    & = & M(r_{\text{max}}).\label{eq: core mass}
\end{eqnarray}

\noindent Solving for the core radius $r_c$ one obtains a closed expression:

\begin{equation}
    r_c = \left[\frac{5040 (2+\beta) M_{\text{core}}}{\left(2^{1/8}-1\right)^{-\frac{3}{2+\beta}} \Gamma\left(\frac{3}{2+\beta}\right)\Gamma\left(8-\frac{3}{2+\beta}\right)}\right]^{1/3}.\label{eq:g0r_c}
\end{equation}

\noindent In the isolated case, where no atmosphere of bosons is around the core like in BEC dark matter collapse simulations, the core mass is also the total mass from the numerical solution $M_{\text{core}} = M(r_{\text{max}})$, and $\beta$ is the only free parameter of $\rho_{\text{core}}(r)$. 
Consequently, $\beta$ is the only fitting parameter to match the ansatz (\ref{eq:soliton_general}) with the numerical solution of the eigenvalue problem.

The optimal $\beta$ value is found via a genetic algorithm using $\textbf{DNA} = (\beta)$ and the fitness function $f(\beta) = \left[\int_{0}^{r_{\text{max}}} |\rho_{\text{core}} - \psi^2| \right]^{-1} dr$, with a tolerance of $10^6$ for the best individual.

We propose an empirical function, similar to (\ref{eq:g0r_c}) but for $g\ne 0$, for the core radius and exponent, subject to the following constraints:

\begin{enumerate}
    \item According to the $\lambda$-scaling, $r_c \sim \rho_c^{-1/4}$ and $r_c \sim g^{1/2}$.
    \item $\beta$ should be $\lambda$-invariant.
    \item When $g = 0, r_c = 1.306\rho_c^{-1/4}$ and $\beta = 0$ are recovered.
\end{enumerate}

\noindent The transformation (\ref{eq:lambda-invariant}) implies that the coefficient $\alpha = g \rho_c^{1/2}$ is $\lambda$-invariant; therefore, $\beta$ should only depend on $\alpha$. Based on these conditions, the following functions are proposed:

\begin{eqnarray}
    r_c & = & 1.306 \rho_c^{-1/4}(1 + a_1 \alpha + a_2 \alpha^2), \label{eq:rca}\\
    \beta & = & b_1 \alpha + b_2 \alpha^2 + b_3 |\alpha|^{1/2}.
    \label{eq:rc}
\end{eqnarray}

\noindent where the fitting parameters are $a_1 = 0.3681 \pm 24.50 \times 10^{-5}$, $a_2 = 0.0905 \pm 31.34 \times 10^{-5}$, $b_1 = 0.2842 \pm 10.71 \times 10^{-5}$, $b_2 = 0.0845 \pm 21.80 \times 10^{-5}$, and $b_3 = -0.0117 \pm 5.443 \times 10^{-5}$.

To see that the general empirical formula is a good approximation to the fundamental state, we show in Figure \ref{fig: rc(g, rhoc)} a plot of $r_c$ as function of $\rho_c$ in Eq. (\ref{eq:rca}) for different values of $g$, both using the solution of the eigenvalue problem and the empirical formula (\ref{eq:g0r_c}). Notice that $r_c$ decreases with increasing $\rho_c$ and increases with increasing $g$ as expected. The plots show how good the empirical formula resembles the properties of the solution to the eigenvalue problem. We also show $\beta$ vs $r_c$ implicit from equations (\ref{eq:rca}-\ref{eq:rc}).  Additionally, $\beta$ changes sign when $g$ changes sign. 

There is an empirical formula in \cite{ChengNiemeyer2021}, similar to (\ref{eq:soliton_general}), constructed based on simulations of core-formation simulations. This formula has two parameters:

\begin{equation}
\rho = \rho_c \left[1 + (2^{1/\beta}-1)\left(\dfrac{r}{r_c}\right)^\alpha\right]^{-\beta}\label{eq:empiricalniemeyer}
\end{equation}

\noindent where the parameters are

\begin{eqnarray}
    \alpha = \alpha_a + (2 - \alpha_a) \tanh^8(\alpha_b (\rho_c g^2)^{-\alpha_c}), \\
    \beta = \beta_a + (2 - \beta_a) \tanh^8(\beta_b (\rho_c g^2)^{-\beta_c}), 
\end{eqnarray}

\noindent and the core radius is more complex:

\begin{equation}
    r_c = 1.308\sqrt{|g|}\left[
    -\frac{a}{2} + \sqrt{\left(\frac{a}{2} \right)^2 + \frac{1}{g^2\rho_0}}
    \right]^{1/2}\nonumber.
\end{equation}
 
\begin{figure*}
    \centering
    \includegraphics[width=12.5cm]{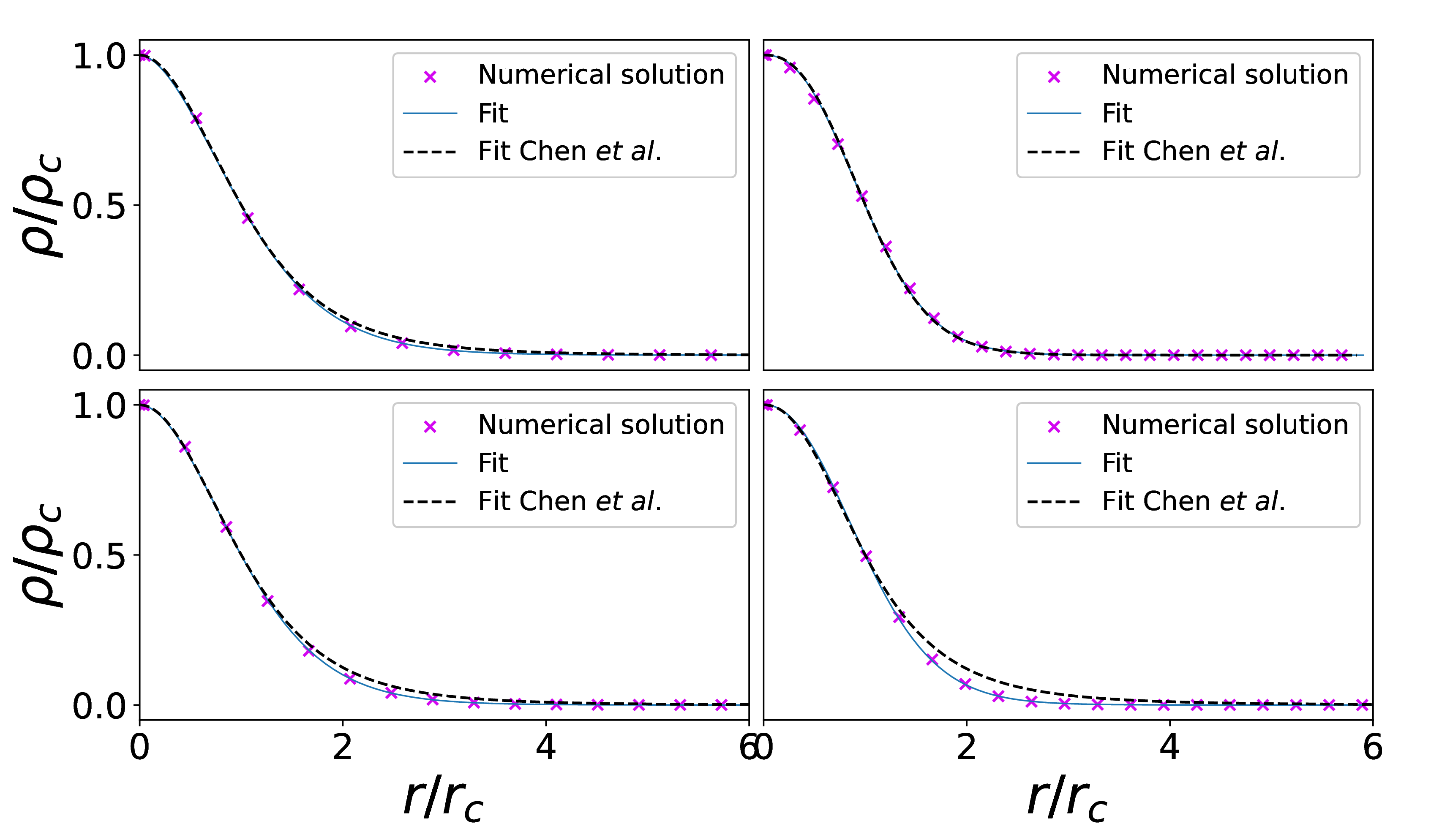}
    \caption{Solutions at the vertices of the parameter space. The top left plot corresponds to $(\psi_c, g) = (2, -0.5)$, the top right plot to $(\psi_c, g) = (2, 0.5)$, the bottom left plot to $(\psi_c, g) = (1, -0.5)$, and the bottom right plot to $(\psi_c, g) = (1, 0.5)$. Points represent the solution to the eigenvalue problem, while continuous lines represent our empirical formula (\ref{eq:soliton_general}) and dashed lines the empirical formula (\ref{eq:empiricalniemeyer}) obtained from simulations in \cite{ChengNiemeyer2021}.}
    \label{fig: comp solutions}
\end{figure*}

\noindent We then compare our profile with this one. Figure \ref{fig: comp solutions} shows the solutions at the corners of our parameter space, at the points $(\psi_c, g) = (1, -0.5), (1, 0.5), (2, -0.5)$, and $(2, 0.5)$. For each of these cases we show with continuous lines the density calculated with the empirical formula (\ref{eq:soliton_general}), with dashed lines the empirical formula found from the core collapse of dark matter in  \cite{ChengNiemeyer2021} and with points the numerical solution, which is the correct solution of the eigenvalue problem. While our formula works fine for these extreme cases of the parameter space, the formula found in \cite{ChengNiemeyer2021} slightly detours from the numerical solution in some cases, probably due to the dynamics that remains in dynamical simulations. These plots show that our results are consistent with previous ones.

\begin{figure*}
    \centering
    \includegraphics[width=6.5cm]{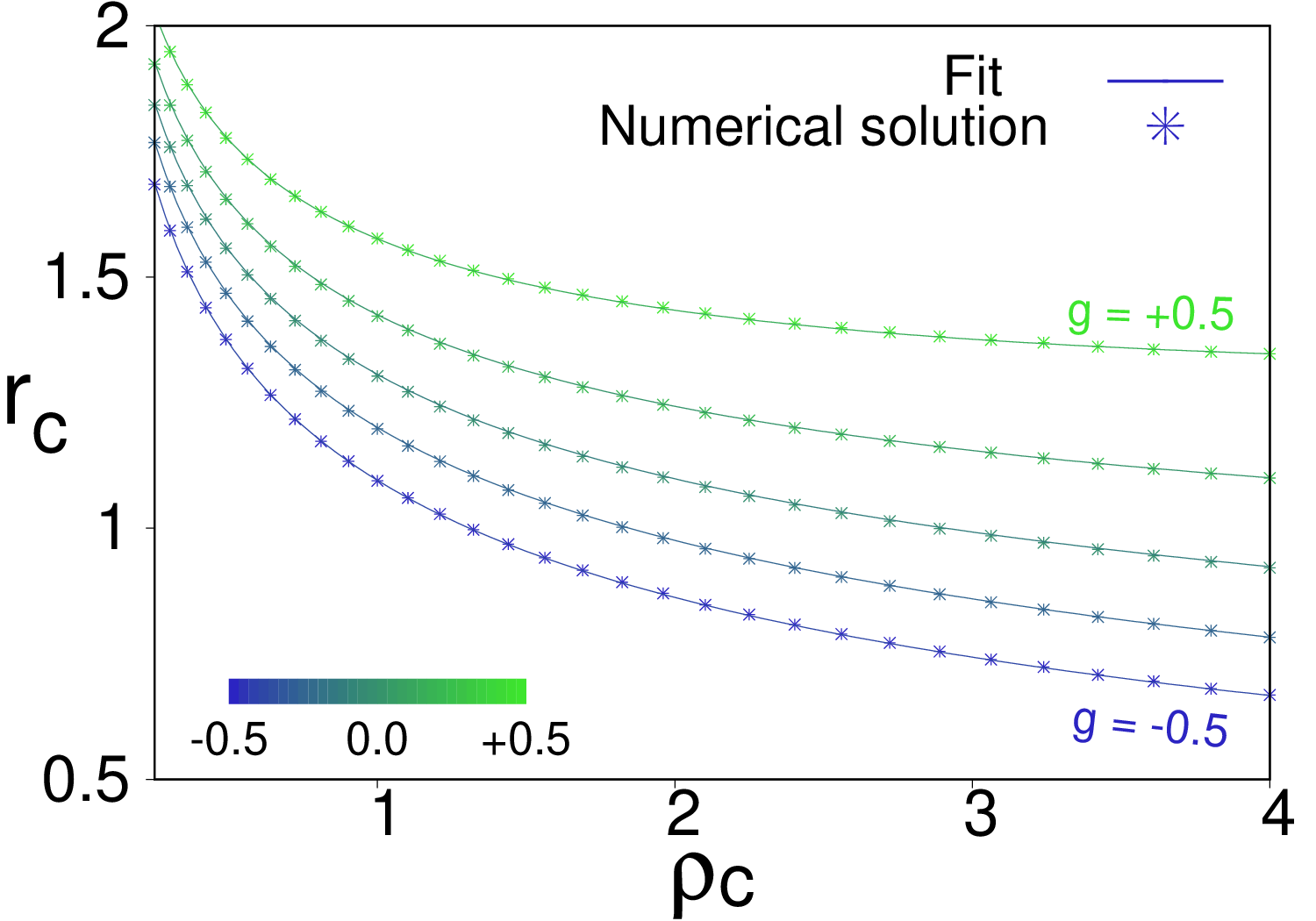}
    \includegraphics[width=6.5cm]{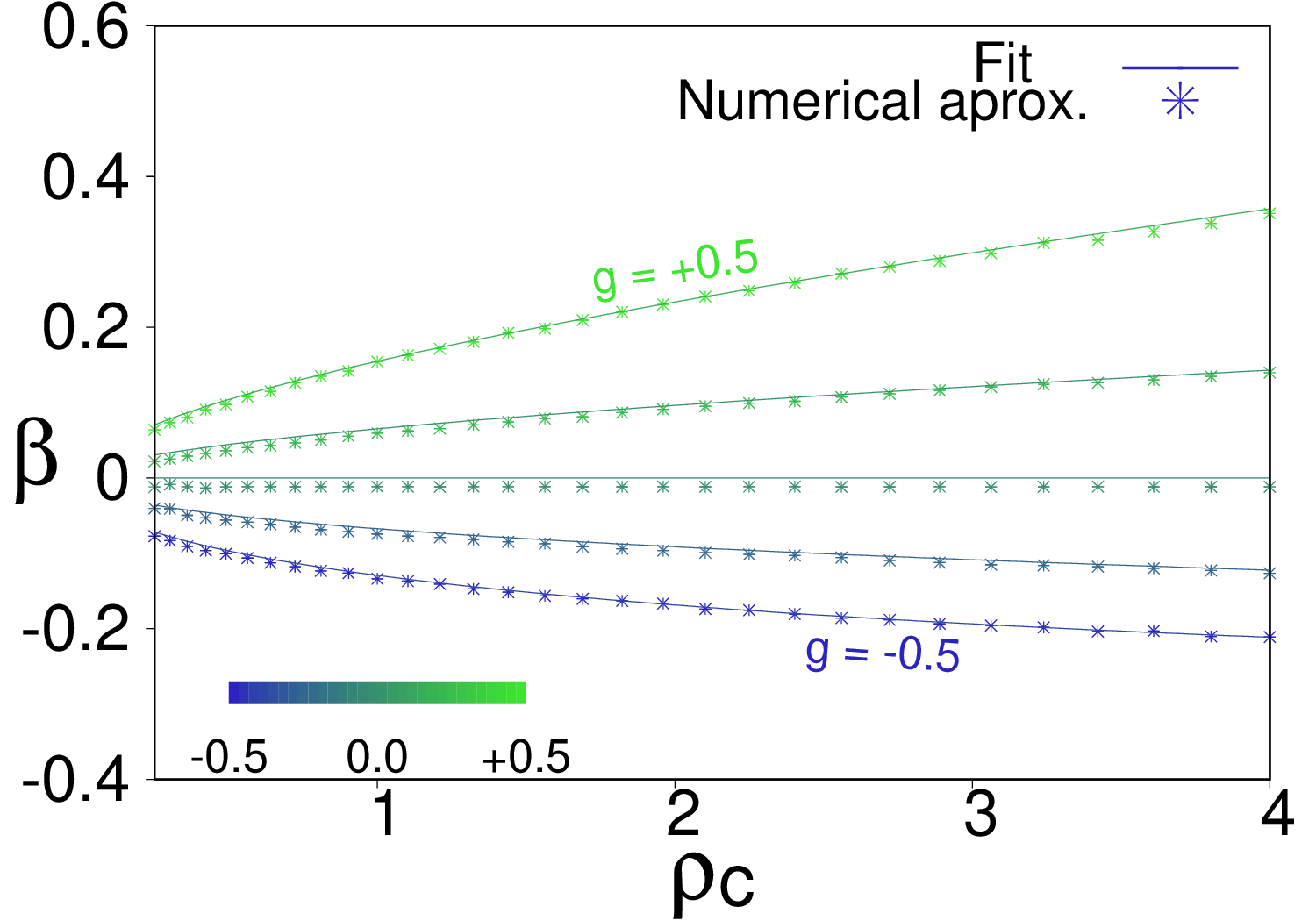}
    \caption{We show the plots of $r_c$ vs $\rho_c$ for the empirical formula and the numerical solution in order to compare them. We also show the parameter $\beta$ as function of $\rho_c$, that indicates how the configuration compacts for different values of $g$.}
    \label{fig: rc(g, rhoc)}
\end{figure*}

\section{Evolution}
\label{sec:evol}

While the solutions found with the GA serve as good approximations to the exact solutions, they are not exact. Let us denote the exact solution of the eigenvalue problem as $\psi_{\text{exact}}$ and the numerical solution as $\psi_{\text{exact}} + \delta\psi$, where $\delta\psi$ represents the error associated with the numerical truncation error of using a discrete domain for its construction. 

Stability can be tested by monitoring the behavior of this error over time when the exact solution is used as the initial condition plus the perturbation. Specifically, we analyze the dynamics triggered by how such perturbation. The stationary solution is deemed stable if the perturbation remains bounded during its evolution, and it is considered unstable if the perturbation grows over time. This error analysis is commonly used to test convergence of numerical solutions of stable solutions \cite{GuzmanUrena2006}, and to test how the errors converge to zero while numerical resolution is increased.

\begin{figure*}
    \centering
    \includegraphics[width=12.5cm]{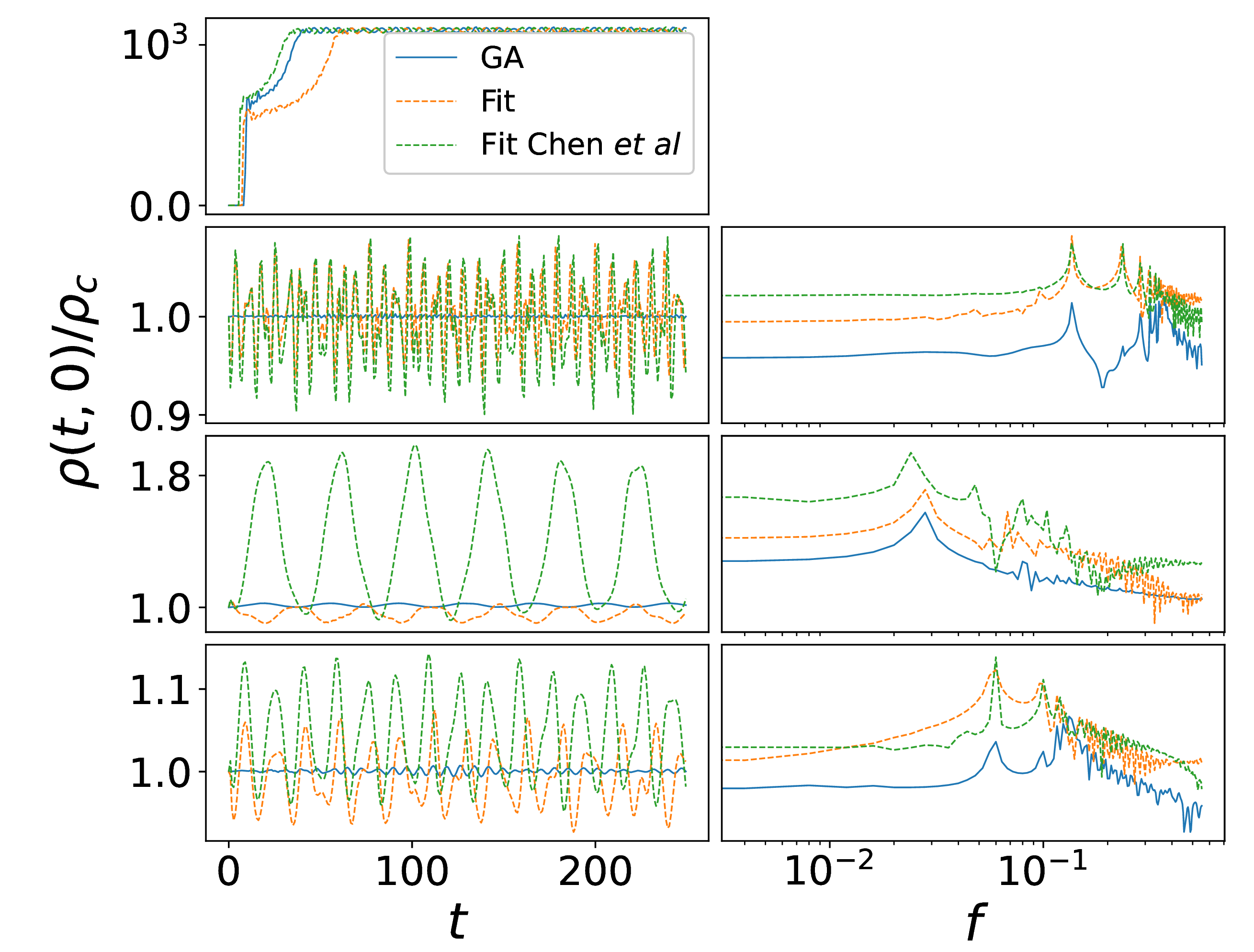}
    \caption{At the left we show the central density as function of time for the evolution of initial conditions obtained from the eigenvalue problem (in blue) and those using the empirical formula (\ref{eq:soliton_general}) (yellow). We also evolve the configurations constructed in \cite{ChengNiemeyer2021} that we draw in green. At the right we show the Fourier Transform of the time-series at the left. The first row corresponds to the unstable case with $(\psi_c, g) = (2, -0.5)$, whose central density runs away indicating the collapse. The subsequent rows correspond to the stable cases with $(\psi_c, g) = (1, -0.5)$, $(\psi_c, g) = (2, -0.5)$, and $(\psi_c, g) = (2, 0.5)$, respectively.}
    \label{fig: central density}
\end{figure*}

Thus the evolution of the numerical solution of the initial value problem has an error that we show does not run away  (see e.g. \cite{GuzmanUrena2004,GuzmanUrena2006}). But we would like to monitor the error when using the empirical density profile as initial condition and see how it behaves. 

For this, we programmed a code that evolves these initial conditions by solving  the time-dependent system (\ref{eq:GP}-\ref{eq:Poisson}) using spherical symmetry. The solution takes place on the same numerical domain $D^d$ used to solve the eigenvalue problem. The code uses the Method of lines to solve the GP equation (\ref{eq:GP}) with a fourth order Runge-Kutta integrator and second order accurate stencils for spatial derivatives. At the origin the  order parameter is extrapolated with a second order accurate approximation. Simultaneously, at each intermediate step of the Runge-Kutta we solve Poisson equation (\ref{eq:Poisson}) outwards from the origin until $r_{max}$.

\begin{figure*}
    \centering
    \includegraphics[width=14.5cm]{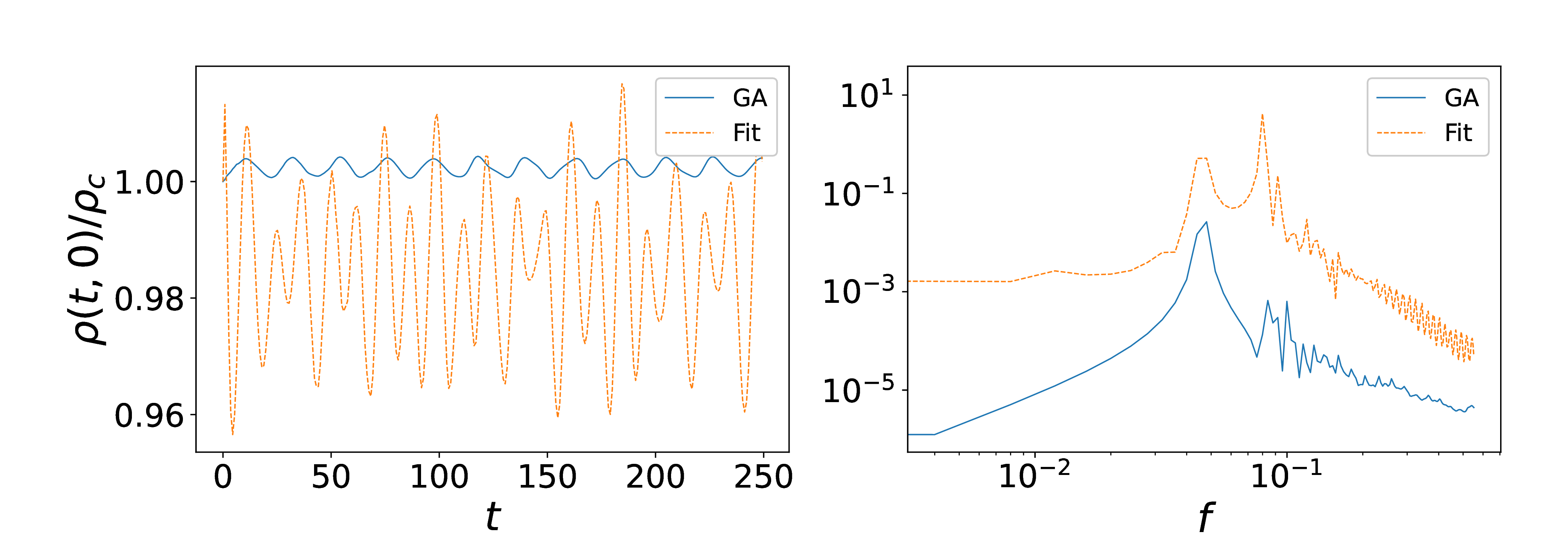}
    \caption{On the left is shown the central density of the evolution of the system using the initial conditions $(\psi_c, g) = (1.0, 0.0)$ for t=250, for the eigensolution and the empirical formula (\ref{eq:soliton_general}). On the right the Fourier Transform that shows that the fundamental frequency is the same for both, the density of the eigensolution and the empirical density profile.}
    \label{fig: central density g =0}
\end{figure*}

\begin{figure}
    \centering
    \includegraphics[width=8cm]{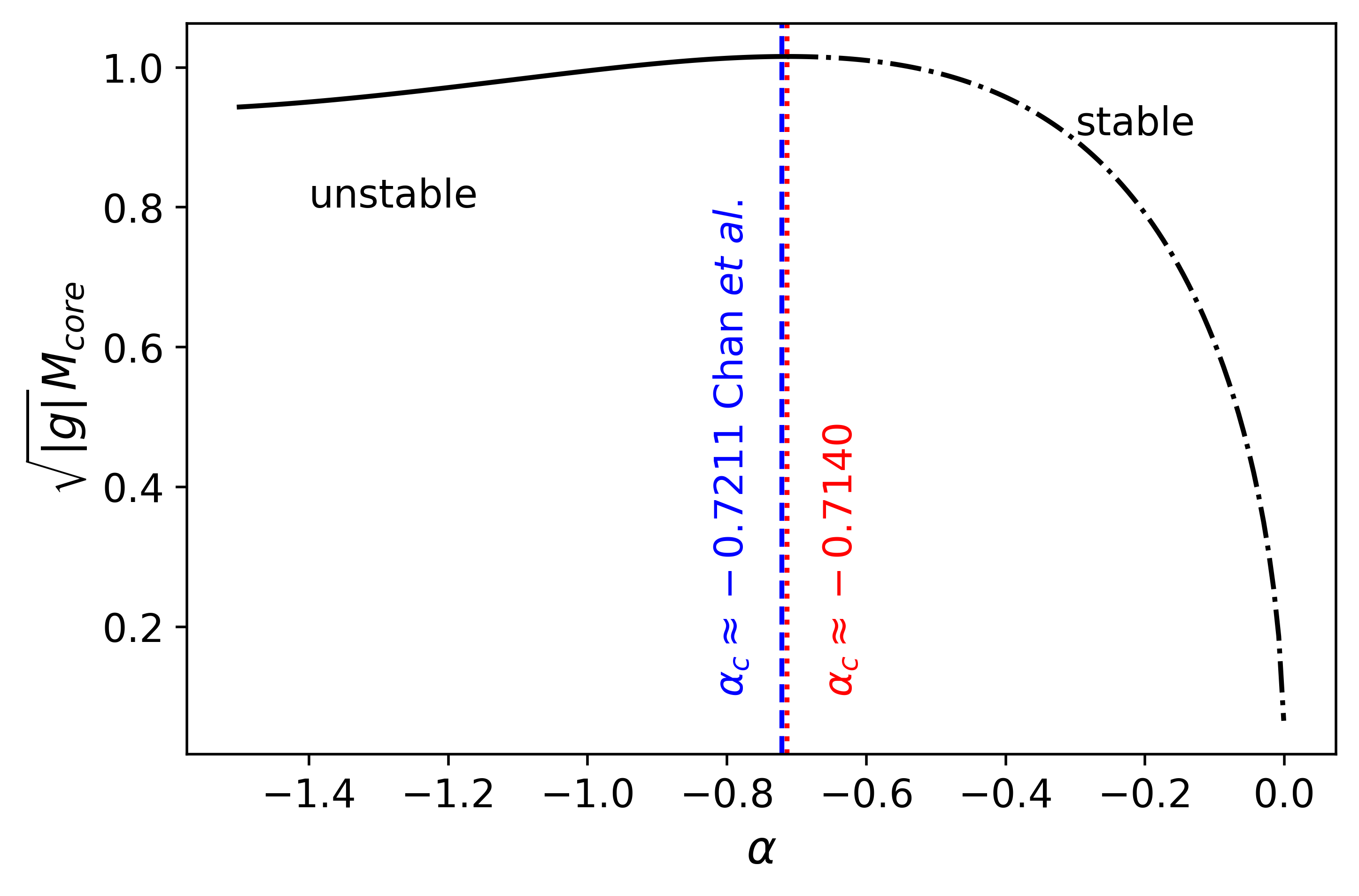}
    \caption{The total mass of the core as a function of the $\lambda$-invariant $\alpha = g \sqrt{\rho_c}$. The vertical red dotted line indicates the critical value $\alpha_c \approx -0.7140$, at which the mass reaches its maximum value. This value divides the stable and unstable branches of solutions. The red dotted line is the result of Chan \textit{et al.}}
    \label{fig:criticalmass}
\end{figure}

The diagnostics we use to monitor the growth of perturbations is the central density at the origin, and the results are shown in Figure \ref{fig: central density}, for the evolution of the solution to the eigenvalue problem and the empirical profile (\ref{eq:soliton_general}) as well as that of formula (\ref{eq:empiricalniemeyer}) obtained from simulations in \cite{ChengNiemeyer2021}, for various combinations of $\psi_c$ and $g$. At the left column we show the time-series of the central density, while the right its Fourier Transform that illustrate the triggered oscillation modes. 

The first case $(\psi_c, g) = (2, -0.5)$ is an unstable solution that collapses due to the attractive nature of self-interaction (negative $g$). In this case the density when initial conditions are the numerical solution blows up at a finite time, whereas when evolving the empirical profile the density also acquires an out of bounds central density, indicating also the instability. In this case the Fourier spectrum says little about the oscillation modes and is not shown.

There are also three stable cases with $(\psi_c, g) = (1, -0.5)$, $(2, \pm 0.5)$, whose central density oscillates with different amplitudes and frequency modes. As expected, the density of the solution of the eigenvalue problem is closer to the exact solution than the empirical formula (\ref{eq:soliton_general}). An implication is that in the first case the amplitude of the oscillations triggered by the truncation error is smaller than in the second case, where the difference between the numerical solution of the eigenproblem and the empirical formula add an extra perturbation. The amplitudes differ approximately by an order of magnitude. Now, what can be seen is that the excited oscillation frequencies coincide and are independent of the oscillation amplitudes.

As a particular case we show the differences between the evolution of the ground state solution and its empirical formula for the case $g=0$ corresponding to FDM. The magnitude of oscillations is particularly important in in this case, because initial conditions involving mergers (Niemeyer varios) and rotation curves (alguno de tula) use cores as workhorse configurations for initial conditions and it is interesting to note how these profiles carry on an intrinsic oscillation. The results are shown in Figure \ref{fig: central density g =0}.

Finally, we carry out an analysis similar to that in \cite{ChengNiemeyer2021}, where a value of $\alpha$ is found that separates the stable and unstable branches of solutions for $g<0$, derived from empirical formula (\ref{eq: core mass}), considering the quantity $\sqrt{|g|} M_{core}$ as a function of the invariant $\alpha$. We show the result in Figure \ref{fig:criticalmass}. The critical value is found where the quantity $\sqrt{|g|}M_{core}$ reaches its maximum value at $\alpha_c \approx -0.7140$, which is similar to the value $\alpha \approx -\sqrt{0.52}$ found by Chan et al. \cite{ChengNiemeyer2021}. This results shows the consistency of our one parameter formula with the formula obtained from simulations of dark matter collapse.

\section{Conclusions}
\label{sec:conclusions}

We have constructed the well known ground state solutions of the GPP system of equations, this time using a Genetic Algorithm. The motivation to implement this type of method, is that it can be used in the case of many parameters, or equivalently a DNA made of coefficients of a multimode wave function. In this sense, this paper is a proof of concept for the usage of this method in core plus halo FDM configurations that we plan to analyze in the near future.

One of the contributions of this work is the construction of a one parameter empirical formula that describes the density profile of ground state solutions with self-interaction. Moreover, this formula works for arbitrary $g$, which is a small step but probably important, with respect to the very general formula for core profiles in \cite{ChengNiemeyer2021} obtained from simulations.

We also evolved the ground state solutions, the density profiles given by our empirical formula, and as a control case also evolved the profiles obtained in \cite{ChengNiemeyer2021}. We found that empirical formulas, even though they produce pretty similar density profiles, their evolution is different. The fact that empirical formulas are an approximate version of the solution of the eigenvalue problem, which is already an approximate solution, produces higher amplitude perturbations. Specifically, we found that the amplitude of the oscillation of stable solutions is bigger than an order of magnitude with respect to those of the ground state solutions. This is relevant because empirical formulas are commonly used as initial conditions for binary and multi mergers of ground state solutions, which can be improved.

Finally, we verified that the evolution of certain configurations with negative selfinteraction can collapse, and found the threshold between stable and unstable solutions using our empirical formula, is consistent with the one found by the analysis in \cite{ChengNiemeyer2021}.

\section*{Acknowledgments}
Carlos Tena receives support within the CONAHCyT graduate scholarship program under the CVU 1303584.
Iv\'an \'Alvarez receives support within the CONAHCyT graduate scholarship program under the CVU 967478. This research is supported by grants CIC-UMSNH-4.9 and CONAHCyT Ciencias de Frontera Grant No. Sinergias/304001. 


\end{document}